\documentclass[prb,showpacs,twocolumn,amsmath,amssymb,floatfix,superscriptaddress]{revtex4-1}

\usepackage{graphicx}
\graphicspath{{Figures/}}

\newcommand{\pdagger}{{\phantom{\dagger}}}

\newcommand{\reff}[1]{Fig.\ \ref{fig:#1}}
\newcommand{\refq}[1]{(\ref{eq:#1})}

\newcommand{\NNcorr}{\langle\hat{\boldsymbol{\sigma}}_i\cdot\hat{\boldsymbol{\sigma}}_j\rangle}
\newcommand{\SiSj}{\langle \mathbf{S}_i\cdot \mathbf{S}_j \rangle}

\usepackage{color}
\newcommand{\NB}[1]{#1}
\begin{document}
	\title{Tunable nanomagnetism in moderately cold fermions on optical lattices}

\author{E.~V.~Gorelik}
\affiliation{Institute of Physics, Johannes Gutenberg University, Mainz, Germany}
\author{N.~Bl\"umer}
\affiliation{Institute of Physics, Johannes Gutenberg University, Mainz, Germany}

\date{\today}
  \begin{abstract}   
Localized defects, unavoidable in real solids, may be simulated in (generically defect-free) cold-atom systems, e.g., via modifications of the optical lattice. We study the Hubbard model on a square lattice with single impurities, pairs of nearby impurities, or lines of impurities using numerically exact determinantal quantum Monte Carlo simulations. In all cases, correlations on the ``impurity'' sites are enhanced either by larger on-site interactions or by a reduced coupling to the environment. 

We find highly nontrivial magnetic correlations, which persist at elevated temperatures and should be accessible in cold-atom systems with current experimental techniques. With improved cooling techniques, these features could be followed towards generic quantum antiferromagnetism in the homogeneous limit. More generally, tunable crossing points between different correlation functions could be used, in a quantum steelyard balance setup, as robust thermometers.
  \end{abstract}
% Hubbard model, SCES, cold gases, Magnetism Hubbard
  \maketitle

%%%%%%%%%%%%%%%%%%%%%%%%%%%%%%%%%%%%%%%%%%%%%%%%%%%%%%%%%%%%%%%%%%%%%%%%%%%%%%%%
%%%%%%%%%%%%%%%%%%%%%%%%%%%%% Introduction %%%%%%%%%%%%%%%%%%%%%%%%%%%%%%%%%%%%%
%%%%%%%%%%%%%%%%%%%%%%%%%%%%%%%%%%%%%%%%%%%%%%%%%%%%%%%%%%%%%%%%%%%%%%%%%%%%%%%%

\section{Introduction}
Imperfections, such as impurity atoms and lattice defects may significantly affect the properties of real materials.\cite{Millis2003} 
This issue is particularly important and complex in the context of strongly correlated materials, where impurity effects may also be employed for detection purposes. 
For example, single impurities were used in the detection of superconducting pairing symmetries within unconventional superconductors\cite{Mackenzie1998, Balastky2006, Alloul2009} and for demonstrating Friedel oscillations.\cite{Sprunger1997}
While cold-atom systems are intrinsically defect-free, impurities can be introduced there in a controlled way, e.g., by employing fine-grained laser speckles,\cite{White2009, Modugno2010, Kondov2013} % NEW ref: Kondov2013
by trapping impurity atoms,\cite{Schirotzek2009, Nascimbene2009, Zipkes2010, Schmid2010} or by projection of (in principle) arbitrary lattice patterns.\cite{Bakr2009}
%Each site can even be manipulated individually using off-resonant laser light or another species of atoms or ions.\cite{Zipkes2010, Targonska2010} 
Even the manipulation of individual sites using off-resonant laser light or another species of atoms or ions has come within reach.\cite{Zipkes2010,Weitenberg2011} % NEW Ref: Weitenberg2011
The unprecedented tunability of artificial impurities provides an exciting route for probing and manipulating the properties of cold atoms, which is attracting increasing theoretical interest for impurity physics in the cold-atom context. 
In particular, the effects of impurities (both static and mobile) on a two-component superfluid Fermi gas in the continuum case as well as trapped in an optical lattice were addressed in detail.\cite{Targonska2010,Vernier2011, Jiang2011, Ohashi2011, Li2012a, Li2012b} 

It is clear that impurities will, in general, affect local and longer-range properties strongly also in the normal phase (with potentially drastic consequences for transport and long-range coherence).
Of particular interest is the effect on magnetism, i.e., spin correlations, in two-flavor fermionic mixtures, as this important aspect of correlation physics is not yet fully under experimental control. 
More specifically, the experimental realization of the long-range antiferromagnetic order that is characteristic of effective single-band Hubbard systems at (or near) half filling, would require further breakthroughs in cooling techniques\cite{Jordens2010}  and, likely, also larger system sizes. 
``Finite-range antiferromagnetism'' should be just within reach of current experiments, when using tunable dimensionality and/or frustration,\cite{Gorelik2012,Chang2013} but requires temperatures low enough for realizing an entropy $s<\ln(2)$ per site throughout the half-filled core of the system. 
This constraint could be relaxed in systems with a small fraction of inequivalent bonds or sites which, loosely speaking,  induce lower-entropy physics locally.\cite{Tarruell2012} 

We suggest to employ this outstanding flexibility of optical lattices to get new insights into the local antiferromagnetism (AF).\cite{Gorelik2012}
As we will show, localized inhomogeneities, in particular impurities, may induce anomalously large spin correlations in the surrounding at or above magnetic ordering temperatures (or the ``spin crossover temperature'' \cite{Paiva2010} in two dimensions), which can be viewed as precursors of  the AF phase. 
This fact, together with the recent progress in experimental techniques allowing precise measurement of double occupancies and of nearest-neighbor (NN) spin correlations in trapped fermionic systems,\cite{Trotzky2010,Greif2011, Greif2013}  % NEW ref: Trotzky2010
could provide the long sought key to the realization and manipulation of quantum magnetism in trapped fermionic systems on optical lattices.\cite{Esslinger2010}

% NEW - start
\NB{In this paper, we consider inhomogeneous variants of the Hubbard model in two dimensions, %in which all sites have only a Hubbard-type on-site interaction $U_i>0$ and are connected by nearest-neighbor hopping matrix elements $t_{ij}$, both of which may vary between ``impurity'' sites and the remaining sites (cf.\ \refq{Hubbard}); 
i.e., we study correlated Anderson-Hubbard-type impurities in a correlated Hubbard-type background, as appears adequate in the cold-atom context.
This is a quite challenging problem already in the one-dimensional case, for which several specialized techniques exist, most notably the Bethe ansatz and the density matrix renormalization group (DMRG) approach. However, also the physics of Hubbard chains is very special so that previous DMRG results for the impurity impact on spin correlations\cite{Costamagna2006} are not directly relevant in higher dimensions.
In the limit of a weakly correlated background and for a single impurity (and uniform hopping), our model reduces to the Wolff model,\cite{Wolff1961} which is closely related to the Anderson impurity model\cite{Anderson1961} and the Kondo model;\cite{Kondo1964} the latter two have been of prime theoretical interest for decades. More recently, various extensions to correlated hosts have been considered, e.g., in the context of high-$T_\text{c}$ superconductivity.\cite{Alloul2009}
%The spin physics of insulating high-$T_\text{c}$ parent compounds can be captured using variants of the Heisenberg model; 
The impact of localized defects on the spin physics of insulating systems (i.e., for a strongly correlated host) in two dimensions has been modelled via vacancies in the Heisenberg model\cite{Martins1997,Song2000,Anfuso2006,Engel2009} and via sites with extra external couplings;\cite{Engel2009} we will relate our work to some of these studies in the following. }
% we will relate our results to some of these studies later.
% NEW - end

The main results of this paper are based on direct determinantal quantum Monte Carlo simulations of the systems of interest, i.e., are exact in the limit of vanishing Trotter discretization and up to statistical errors. In addition, we provide data obtained using a real-space extension (i.e., going beyond the so-called local density approximation) of dynamical mean-field theory (DMFT).\cite{Metzner1989, Georges1996} This comparison helps in identifying the essential physics generating the observed correlation patterns. At the same time, our assessment of the real-space DMFT points out important limitations of this popular approximation.

%%%%%%%%%%%%%%%%%%%%%%%%%%%%%%%%%%%%%%%%%%%%%%%%%%%%%%%%%%%%%%%%%%%%%%%%%%%%%%%%%%%%%%%%%%%%%%%%%%%%
%%%%%%%%%%%%%%%%%%%%%%%%%%%%%%%%%%%%%%%%%%%%%%%%%%%%%%%%%%%%%%%%%%%%%%%%%%%%%%%%%%%%%%%%%%%%%%%%%%%%
\section{Model and methods}
We consider the single-band Hubbard Hamiltonian
\begin{equation}\label{eq:Hubbard}
  \hat{H} =\! - \sum_{\langle ij\rangle ,\sigma}  t_{ij}\, \hat{c}^{\dag}_{i\sigma}
\hat{c}^\pdagger_{j\sigma} 
+   \sum_{i}  U_i\, \left(\hat{n}_{i\uparrow}-\tfrac{1}{2}\right)\left( \hat{n}_{i\downarrow}-\tfrac{1}{2} \right)
\end{equation}
with (in general) bond-dependent hopping amplitudes $t_{ij}$ between nearest neighbors and (in general) site-specific local interactions $U_i$; $\hat{n}_{i\sigma}=\hat{c}^{\dag}_{i\sigma}\hat{c}^\pdagger_{i\sigma}$ denote spin-resolved densities.
Particle-hole symmetry (at zero chemical potential) guaranties half filling at each site: $\langle \hat{n}_{i\uparrow} +  \hat{n}_{i\downarrow} \rangle = 1$, thereby avoiding the sign problem in determinantal quantum Monte Carlo calculations.

In the homogeneous case ($t_{ij} = t$, $U_i=U$) and on a simple cubic lattice, the model \refq{Hubbard} exhibits long-range antiferromagnetic (AF) order at low temperatures $T<T_\text{N}$, corresponding to a critical entropy per particle $s_\text{N} \lesssim 0.34 \approx \ln(2)/2$.\cite{Gorelik2012} 
Even at elevated temperatures, strong AF correlations remain dominant in the paramagnetic phase, roughly up to 
the mean-field critical entropy $s_\text{N}^\text{DMFT}\stackrel{U\to\infty}{\longrightarrow} \ln(2)\approx 0.69$; later, we will denote the corresponding spin-crossover temperature as $T_{\text{sc}}$.
This ``finite-range antiferromagnetism'' has remarkably universal characteristics: local properties and short-range correlations depend only weakly on the dimensionality at constant $s$, which makes its realization a worthwhile and realistic goal for cold-atom experiments.\cite{Gorelik2012}

The same can be expected for the nanomagnetic properties of interest in this paper, i.e., the (change of) AF correlations induced by impurities embedded in otherwise homogeneous systems: the essential physics and mechanisms should be very similar in two and three dimensional host systems (at constant $s$ and for equivalent interaction strengths). 
We will consider square lattices in this paper, which are not only easier to simulate numerically, but also offer important advantages (in particular, the possibility of single-site addressing) in cold-atom experiments. 

We examine two types of local inhomogeneities, both tailored towards enhanced correlations. 
In the first case, we introduce deviating on-site interactions $U_\text{imp}=U' > U$ on some of the lattice sites (compared to the homogeneous background with interaction $U$), whereas both the nearest-neighbor hopping amplitudes $t_{ij}=t$ and the filling are kept constant across the sample. 
The on-site interaction on the ``impurity'' sites $U'=8t=2U$ was chosen to maximize the spin-crossover temperature $T_\text{sc}(U)$, which reaches a peak value of $T_\text{sc}^\text{imp}\approx 0.4t$ at the intermediate coupling $U=8t$.\cite{Paiva2010,Gorelik2012}  In contrast, the weak coupling $U=4t$ chosen for the remaining sample is associated with a much smaller $T_\text{sc}^\text{hom}\approx 0.2t$, so that we can expect to see strong impurity effects at the moderately low temperature $T=0.25 t$ (as $T_\text{sc}^\text{hom} < T < T_\text{sc}^\text{imp}$).

The second type of ``impurities'' we are going to consider is realized by a reduced coupling to the environment: all the hopping amplitudes $t_{\text{imp},j}=t'$ between such impurity and its nearest neighbors are smaller than the NN hopping amplitude $t$ throughout the rest of the sample, whereas the on-site interaction is constant for all the lattice sites. 

If not explicitly stated otherwise, results presented in the following were obtained employing determinantal quantum Monte Carlo (DQMC)\cite{Blankenbecler1981, QUEST_link,Rost2012} with finite Trotter discretization $\Delta\tau \le 0.1/t$ and up to 100 sites. 

%%%%%%%%%%%%%%%%%%%%%%%%%%%%%%%%%%%%%%%%%%%%%%%%%%%%%%%%%%%%%%%%%%%%%%%%%%%%%%%%%%%%%%%%%%%%%%%%%%%%
%%%%%%%%%%%%%%%%%%%%%%%%%%%%%%%%%%%%%%%%%%%%%%%%%%%%%%%%%%%%%%%%%%%%%%%%%%%%%%%%%%%%%%%%%%%%%%%%%%%%
\section{Results}
In the following, we discuss effects of impurities on local and short-range magnetic properties that could be relevant for cold fermionic gases on optical lattices. 
We focus on two observables that are accessible in cold-atom experiments and have a direct connection with antiferromagnetism: the double occupancy\cite{Gorelik2010, Gorelik2012, Jordens2008} $D=\langle \hat{n}_{i\uparrow} \hat{n}_{i\downarrow} \rangle$ and the nearest-neighbor (NN) spin correlations $\SiSj$.\cite{SiSj,Trotzky2010,Greif2011, Greif2013} % NEW ref: Trotzky2010

\subsection{Impurities with increased local interaction}
Let us first take a look at the impact of locally enhanced on-site repulsions, as depicted in \reff{Uimp_map_DQMC}. 
We consider four different implanted impurity patterns, marked by bright white dots: a single impurity site ($1^\textrm{st}$ column of \reff{Uimp_map_DQMC}), a pair of impurities on neighboring sites ($2^\textrm{nd}$ column) and two patterns with next-nearest neighbor pairs (in the taxi-cab metric) along the axis and diagonal ($3^\textrm{rd}$ and $4^\textrm{th}$ columns, respectively). 
%%%%%%%%%%%%%%%%%%%%%%%%%%%%%%%%%%%%%%%%%%%%%%%%%%%%%%%%%%%%%%%%%%%%%%%%%%%%%%%%%
\begin{figure}[t] %FIG.1%
\includegraphics[width=\columnwidth]{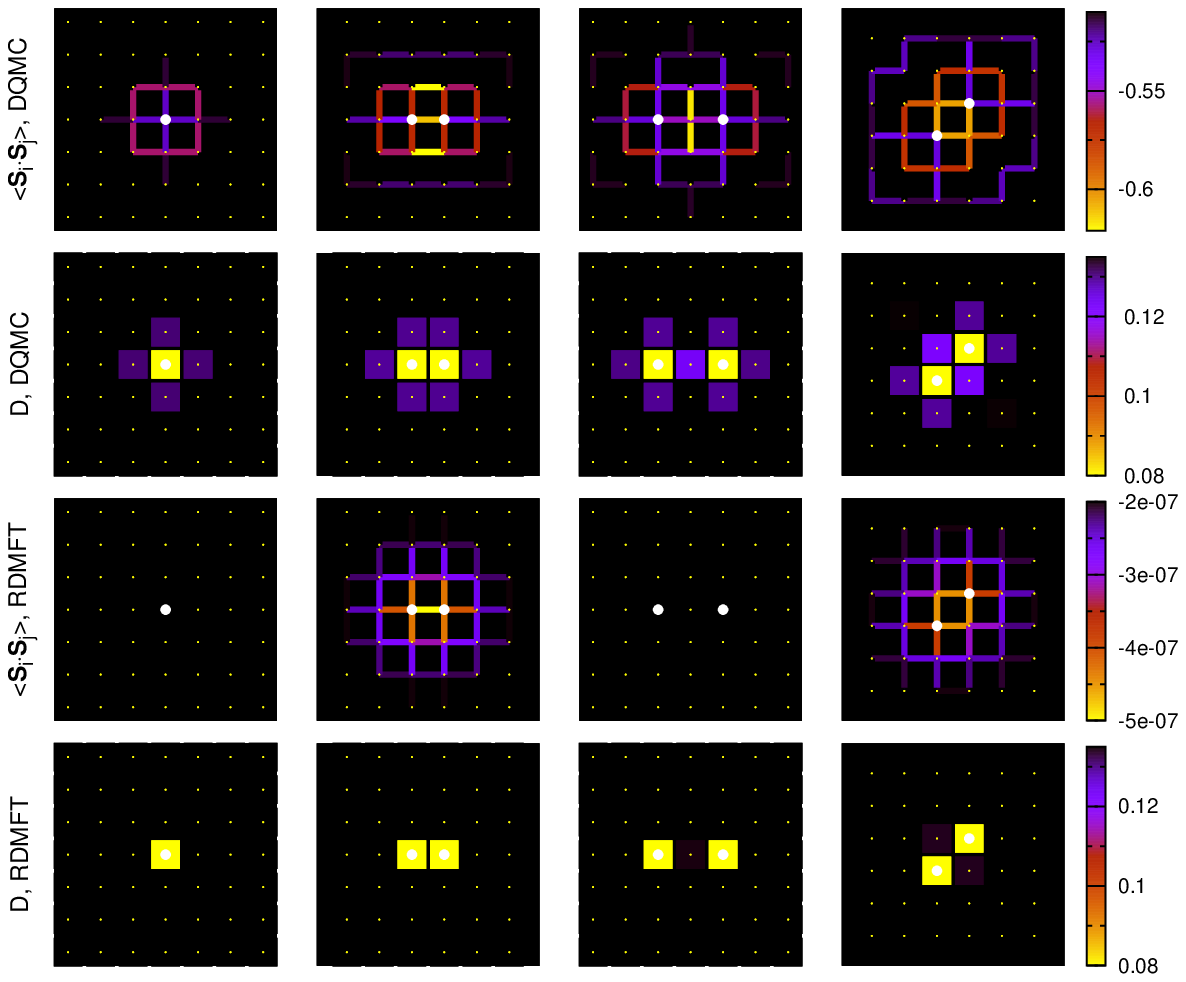}
\caption{(Color online)
The effect of single or paired impurity sites with on-site interaction $U'=8t$ introduced into $U=4t$ medium at $T=0.25t$. 
Nearest-neighbor spin correlations $\SiSj$ ($1^\textrm{st}$ row) and double occupancy $D$ ($2^\textrm{nd}$ row) from DQMC calculations as compared to $\SiSj$ ($3^\textrm{rd}$ row) and $D$ ($4^\textrm{th}$ row) from RDMFT calculations. 
Impurity sites are marked with bright white dots.
}\label{fig:Uimp_map_DQMC}
\end{figure}
%%%%%%%%%%%%%%%%%%%%%%%%%%%%%%%%%%%%%%%%%%%%%%%%%%%%%%%%%%%%%%%%%%%%%%%%%%%%%%%%%
Due to its local character, the changes in the double occupancy closely follow the implanted impurity pattern: the second row of \reff{Uimp_map_DQMC} demonstrates a local suppression of the double occupancy by about $30\%$ at the sites with doubled on-site interaction, reflecting strongly enhanced on-site correlations. 
By proximity effects, the double occupancy is also slightly reduced, by about $5-10\%$ on the immediate neighbors of each impurity.
The nearest-neighbor spin correlations ($1^\textrm{st}$ row of \reff{Uimp_map_DQMC}) are also, in general, enhanced in the vicinity of the impurities. 
However, this effect is by no means restricted to the bonds extending from the impurity sites. 
Already in the case of a single isolated impurity site (the first column of \reff{Uimp_map_DQMC}), the strongest spin correlations appear along bonds between nearest and the next-nearest neighbors of the site with doubled interactions, forming a square around the impurity; here, the enhancement of $\SiSj$ is more than twice as large as for the bonds extending from the impurity. 
% NEW - start
\NB{Note that this phenomenology goes beyond the resonant-valence-bond (RVB) picture proposed for vacancies in the Heisenberg model,\cite{Martins1997} where the elimination of some spin correlations (between the ``vacancy'' and the enviroment) enhances correlations at adjacent bonds; in the present case, such an RVB-type explanation cannot be applied since all NN spin correlations involving the NN sites of the impurity are enhanced.}
% NEW - end

In the case of multiple impurities, the patterns become much more complex: in the case of a NN impurity pair ($2^\textrm{nd}$ column), the bonds above and below the impurity pair are particularly strongly correlated; with one extra spacing ($3^\textrm{rd}$ row), the central bonds orthogonal to the impurity pair become enhanced. 
We will shed more light on this fascinating physics in the following.

Apart from the intrinsic interest, the above systems can also serve as extreme test cases for the applicability and accuracy of the real-space extension of dynamical mean-field theory (RDMFT)\cite{Helmes2008, Snoek2008, Gorelik2010, Gorelik2011, Bluemer11CCP}, here evaluated using a Hirsch-Fye quantum Monte Carlo impurity solver.\cite{Hirsch86,Bluemer07,Bluemer2013}
As expected, RDMFT yields rather accurate estimates of local observables such as the double occupancy\cite{Gorelik2012} ($4^\textrm{th}$ row of \reff{Uimp_map_DQMC}) on the impurity sites
(while the weaker proximity effects observed in the results of DQMC calculations are nearly lost).
However, RDMFT neglects nonlocal correlations arising from fluctuations: 
$\SiSj \stackrel{\text{DMFT}}{\longrightarrow} \langle \mathbf{S}_i \rangle \cdot \langle \mathbf{S}_j \rangle$. 
Consequently, nontrivial (i.e., nonlocal) spin correlations vanish in the absence of static magnetization. 
As the bulk parameters are outside the mean-field AF phase, the DMFT estimates of $\SiSj$ ($3^\textrm{rd}$ row of \reff{Uimp_map_DQMC}) vanish exactly in the bulk; 
inserting a single impurity or an impurity pair with sufficient spacing does not change the situation, as seen in the first and third column of \reff{Uimp_map_DQMC}. 
However, the finite (small) spin correlations in the second and fourth column of \reff{Uimp_map_DQMC} indicate that already a single impurity pair can give rise to a local magnetization pattern if the temperature is not too far above the bulk mean-field Neel temperature.
The corresponding mean-field spin correlations, however, fail to reproduce the DQMC data both quantitatively and qualitatively, as is clear from the comparison of the $1^\textrm{st}$ and $3^\textrm{rd}$ rows of \reff{Uimp_map_DQMC}.

% NEW start
\NB{Moreover, the true correlation patterns at finite interactions (first row of \reff{Uimp_map_DQMC}) are also substantially different from those previously observed\cite{Anfuso2006} in the case of two vacancies in the Heisenberg model (to which our model maps, e.g., in the limit $U'\gg U\gg t$), in analogous configurations.}
% NEW end
These comparisons reemphasize that the spatial structure of the impurity-induced spin correlations observed in the DQMC data is by far not trivial.

Still, one might ask whether the patterns seen in the case of multiple impurities (in the first row of \reff{Uimp_map_DQMC}) can be understood on the basis of the single-impurity effects. More precisely: are the changes in the NN spin correlations induced by the impurities mere superpositions of the effects of the individual impurities or do we see genuine many-impurity effects? In order to answer this question, we have computed the differences $\Delta\SiSj$ between the NN spin correlations in the presence of impurities and those of the homogeneous system. For the special case of a single impurity, we denote the result as $\Delta\SiSj_0$, i.e. with an index 0; it is shown in the top left corner of \reff{sum1imp}. 
%%%%%%%%%%%%%%%%%%%%%%%%%%%%%%%%%%%%%%%%%%%%%%%%%%%%%%%%%%%%%%%%%%%%%%%%%%%%%%%%
\begin{figure}[t] %FIG.2%
\includegraphics[width=\columnwidth]{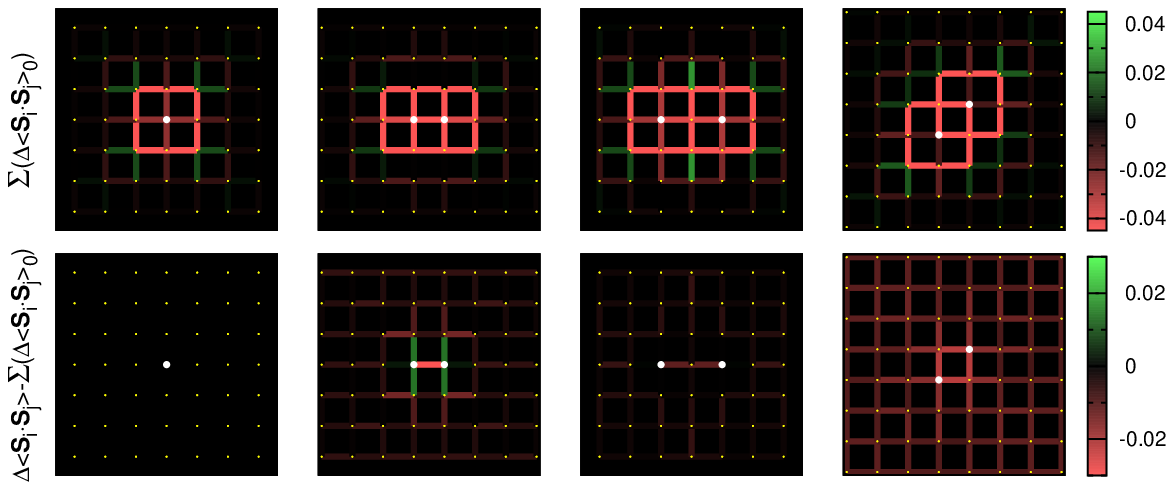}
\caption{(Color online)
First row: the sum of the effects of two single impurities $\sum\left(\Delta\SiSj_0\right)$.
For the ease of comparison we consider the change in NN spin correlations with respect to the homogeneous impurity-free value $\SiSj_\text{hom}$: $\Delta\SiSj=\SiSj - \SiSj_\text{hom}$.
Red/green colors indicate enhanced/suppressed spin correlations (relative to the impurity-free homogeneous value).
Second row: difference of $\Delta\SiSj$ and $\sum\left(\Delta\SiSj_0\right)$.
All parameters are as in \reff{Uimp_map_DQMC}.
Impurity sites are marked with bright white dots.
}\label{fig:sum1imp}
\end{figure}
%%%%%%%%%%%%%%%%%%%%%%%%%%%%%%%%%%%%%%%%%%%%%%%%%%%%%%%%%%%%%%%%%%%%%%%%%%%%%%%%%
The other subfigures in the first row of \reff{sum1imp} have been obtained by adding up these single-impurity results for each of the impurities (bright dots). 
Indeed, these superpositions look remarkably similar to the patterns seen in the first row of \reff{Uimp_map_DQMC} (and \reff{Uhopp_map_DQMC}); if plotted using the same color code, both data sets would hardly be distinguishable from each other. 
If we, however, subtract the exact results from the superpositions, as shown in the second row of \reff{sum1imp}, we see that a pair of NN impurities (second column) induces singlet-type physics, i.e. a stronger correlation between the impurities and reduced correlations between the impurities and the environment (while this observable vanishes by definition in the single-impurity case). 
This many-impurity effect decays rapidly when the distance is doubled (third column); the slight overall enhancement in the case of a diagonal pair (fourth column) might be an artifact due to finite-size effects in the underlying simulations.
We can conclude that, for interaction type impurities, intrinsic many-impurity effects are significant only when these are directly connected by hopping bonds.

\subsection{Impurities with reduced coupling to the bulk}
A direct experimental realization of the impurity type considered above might be difficult, since it requires local changes of the on-site interaction together with adjustments of the chemical potential on the impurity sites to keep the density at half filling. 
An alternative strategy for inducing stronger local correlations at selected ``impurity sites'', namely a partial decoupling by reduced hopping amplitudes between the impurity sites and their environment, seems to be more feasible.\cite{Greif2013}

In the following, we choose $t_{\text{imp},j}=t'=0.5t$, keeping all interactions the same as for the environment (with $U=4t$).
With this choice, the ratio $U/t'$ between interaction and hopping on the impurity sites is the same as in the previous section (i.e., $U/t'= U'/t=8$). 
Note that the ratios between temperature $T$ and hopping ($t$ or $t'$) or interaction ($U'$ or $U$) deviate by a factor of two in both systems. 
%%%%%%%%%%%%%%%%%%%%%%%%%%%%%%%%%%%%%%%%%%%%%%%%%%%%%%%%%%%%%%%%%%%%%%%%%%%%%%%%
\begin{figure}[t] %FIG.3%
\includegraphics[width=\columnwidth]{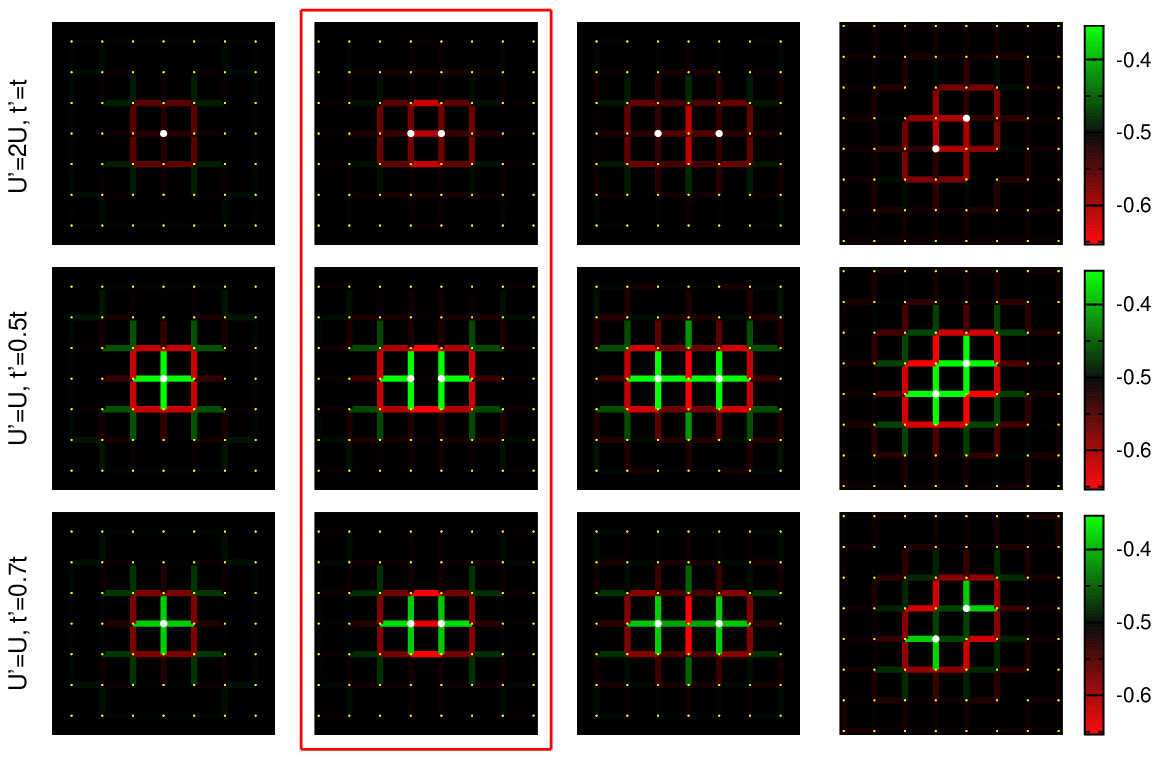}
\caption{(Color online)
The effect of impurity sites introduced into a $U=4t$ medium at $T/t=0.25$ on nearest-neighbor spin correlations $\SiSj$. 
First row: $U'=8t=2U$, $t'=t$; 
second row:  $U'=4t=U$, $t'=0.5t$; 
third row:  $U'=4t=U$, $t'=0.7t$. 
Red/green colors indicate enhanced/suppressed spin correlations (relative to the impurity-free homogeneous value).
A red frame highlights the system with two impurities on neighboring lattice sites discussed in detail in the main text.
Impurity sites are marked with bright white dots.
}\label{fig:Uhopp_map_DQMC}
\end{figure}
%%%%%%%%%%%%%%%%%%%%%%%%%%%%%%%%%%%%%%%%%%%%%%%%%%%%%%%%%%%%%%%%%%%%%%%%%%%%%%%%%
For our particular choice of parameters, however, the impact of this discrepancy should be minimal, as the magnitude of NN spin correlations in homogeneous systems with $U/t=4$ at $T/t=0.25$ and with $U/t=8$ at $T/t=0.5$ are almost identical [cf.\ \reff{nn_vs_T}(a)].

As seen in the second row of \reff{Uhopp_map_DQMC}, the effect of the reduced hopping on spin correlations is dramatic and differs much from the previously discussed case of impurities with increased local interactions, shown again in the first row of \reff{Uhopp_map_DQMC} for comparison. Most notably,
we observe a strong suppression of spin correlations between the impurities and their ``normal'' NN 
(for $t'=0.5t$, shown in the second row in \reff{Uhopp_map_DQMC}, by approximately a factor of two), 
whereas NN spin correlations between the sites surrounding the impurities are significantly amplified. 
In contrast, the spin correlation between NN impurities ($2^\textrm{nd}$ column of \reff{Uhopp_map_DQMC}) is enhanced at $t'=0.7t$ ($3^\textrm{rd}$ row), while it recovers the background value at $t'=0.5t$ ($2^\textrm{nd}$ row). It is clear that such behavior cannot arise from overlaying single-impurity effects (in contrast to the case of interaction type impurities, cf.\ \reff{sum1imp}). In other words: all patterns depend crucially on the precise locations of all impurities (and also on the tuning parameter $t'$). 
% NEW start
\NB{As in the case of interaction-type impurities, our DQMC results for the Hubbard model do not share obvious similarities to the correlation patterns observed\cite{Anfuso2006} in the case of two vacancies in the Heisenberg model in analogous configurations.}
% NEW end

%%%%%%%%%%%%%%%%%%%%%%%%%%%%%%%%%%%%%%%%%%%%%%%%%%%%%%%%%%%%%%%%%%%%%%%%%%%%%%%%%
\begin{figure}[t] %FIG.4%
\includegraphics[width=\columnwidth]{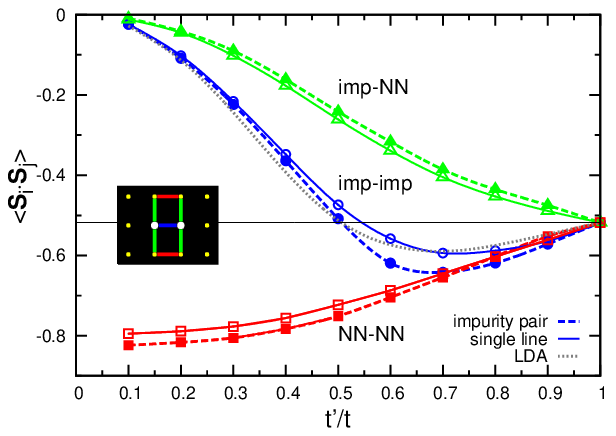}
\caption{(Color online) 
The effect of impurity sites with $U'=4t=U$, $t_{\text{imp},j}=t'$ introduced into a medium with $U=4t$ at $T/t=0.25$ on NN spin correlations: 
$\SiSj$ as a function of $t'$ for a system with two impurities on the neighboring lattice sites as in the second column of \reff{Uhopp_map_DQMC} (dashed lines and open symbols) or single line of impurities (solid lines and filled symbols). 
The inset illustrates the choice of the bonds $i-j$ for the case of an impurity pair. 
Black solid line: NN spin correlations $\SiSj_\text{hom}$ for the homogeneous system with $U/t=4$, $T/t=0.25$.
Dotted line: NN spin correlations for a homogeneous system with the parameters of the impurity (LDA), $\SiSj_\text{LDA}$.
}\label{fig:nn_vs_timp}
\end{figure}
%%%%%%%%%%%%%%%%%%%%%%%%%%%%%%%%%%%%%%%%%%%%%%%%%%%%%%%%%%%%%%%%%%%%%%%%%%%%%%%%%
Let us focus now on the case of two impurities on neighboring lattice sites, for which the full space distribution of NN spin correlations is shown in the second column of \reff{Uhopp_map_DQMC}.
NN spin correlations relate to a pair of neighboring lattice sites, which we are going to refer to as a ``bond'' $i-j$ between lattice sites $i$ and $j$.
For the following quantitative analysis,  we choose three types of bonds that capture the main features of the emerging pattern, as illustrated in the inset of \reff{nn_vs_timp}:
the one between impurity sites (imp-imp, shown in blue on the inset of \reff{nn_vs_timp}), two equivalent bonds between ``normal'' sites next to impurities parallel to impurity pair (NN-NN, shown in red on the inset of \reff{nn_vs_timp}), and four equivalent bonds (orthogonal to imp-imp) between each impurity and the neighboring ``normal'' sites (imp-NN, shown in green on the inset of \reff{nn_vs_timp}).
These three types of bonds demonstrate very different dependencies of spin correlations on $t'/t$.
AF spin correlations along imp-NN bonds (filled triangles in \reff{nn_vs_timp}) are suppressed by the reduction of $t'/t$, and converge to zero as $t'$ decreases to zero.
In contrast, a reduction of $t'/t$ induces amplified AF correlations along NN-NN bonds (filled squares), due to the reduced coordination of the involved sites.
Spin correlations imp-imp between impurity sites (filled circles) show more complex behaviour upon variation of $t'/t$ with a slight {\it increase} of the correlations amplitude upon tuning $t'/t$ down from $1.0$ to $0.7$, and subsequent {\it decrease} with further lowering of $t'/t$.
At $t'/t\approx 0.5$ imp-imp correlations are matching again the homogeneous impurity-free value $\SiSj_\text{hom}$ (black solid line in \reff{nn_vs_timp}).
Similar trends are observed in $\SiSj_\text{LDA} (t'/t)$ for a homogeneous system with the parameters of the impurity serving here as a local density approximation (LDA) for the initial system (dotted line in \reff{nn_vs_timp}).
This shows that the physics of the full system (two NN impurities) can be understood, to a first approximation, in a local LDA-type picture. Note, however, that the enhancement of imp-imp AF correlations (filled circles) at $t'/t\approx 0.7$ is about twice as strong as the LDA prediction; i.e., singlet physics or, more generally, reduced dimensionality must play an important role.

In order to answer this question and to allow for comparisons also with the correlations involving normal sites (imp-NN and NN-NN), we consider a system containing a single line of impurities, i.e., a single layer of sites with reduced hopping, sandwiched between multiple layers of ``normal'' sites (with periodic boundary conditions); corresponding results are shown using open symbols (and solid lines) in \reff{nn_vs_timp}. In this case, imp-imp refers to NN correlations within the ``impurity'' layer, while NN-NN denotes NN correlations within either of the adjacent layers and imp-NN refers to correlations across the layers. 
We find that the imp-imp correlation in the stacked system (open circles), despite its reduced dimensionality, is very close to the LDA prediction (dotted lines), at least in the parameter region $t'/t\approx 0.7$. This shows that the enhancement seen for the impurity pair is really an effect of singlet physics, i.e., specific for NN impurity pairs. As we will show later, this nonlocal effect arises only at quite low temperatures.
In comparison, the other correlation functions differ less between the two systems (impurity pair versus stacked); however, the deviations from the homogeneous limit (t'/t=1) are generically stronger for the (zero-dimensional) impurity pair system (except for imp-imp at $t'/t\lesssim 0.5$).

Obviously, the temperature $T/t=0.25$ chosen so far was sufficiently low for realizing interesting spin correlation patterns.
It is important to check whether this remains true towards elevated temperatures which are more easily realized in experiments. 
The evolution of impurity-induced changes in $\SiSj$ with temperature is illustrated in \reff{imp_map_vs_B}.
%%%%%%%%%%%%%%%%%%%%%%%%%%%%%%%%%%%%%%%%%%%%%%%%%%%%%%%%%%%%%%%%%%%%%%%%%%%%%%%%%
\begin{figure}[t] %FIG.5%
\includegraphics[width=\columnwidth]{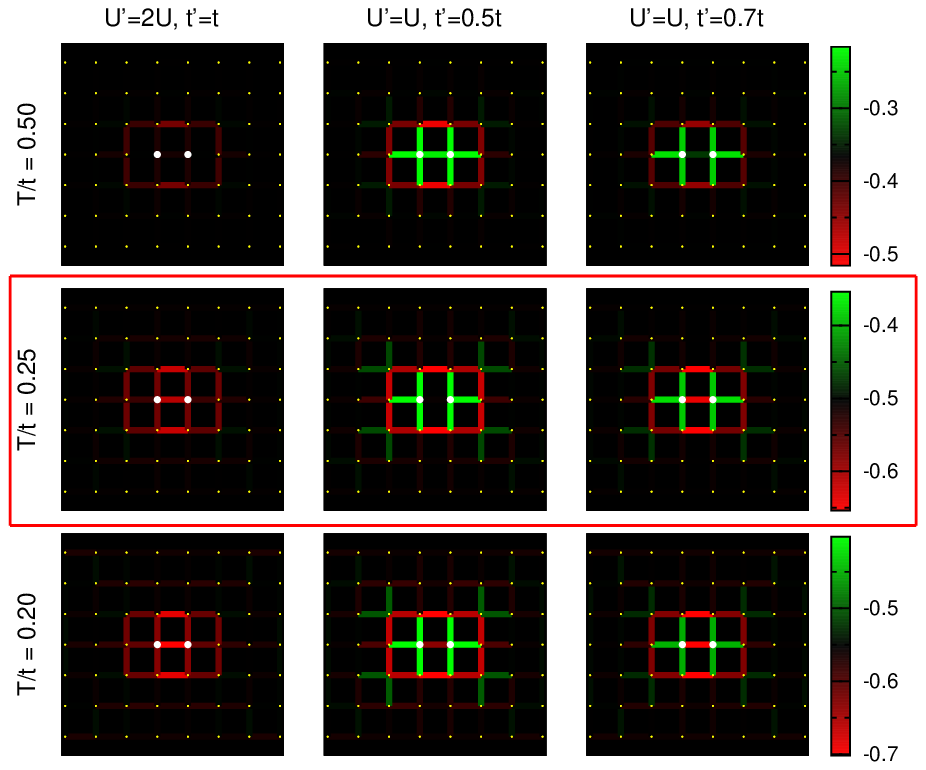}
\caption{(Color online) 
Temperature dependence of impurity effects on NN spin correlations $\SiSj$. 
First column: $U'=8t=2U$, $t'=t$; 
second column:  $U'=4t=U$, $t'=0.5t$; 
third column:  $U'=4t=U$, $t'=0.7t$. 
Amplified spin correlations (relative to the impurity-free homogeneous value for specific $T/t$) are shown in red, suppressed ones in green.
The data in the red frame is repeated from \reff{Uhopp_map_DQMC}.
Impurity sites are marked with bright white dots. 
}\label{fig:imp_map_vs_B}
\end{figure}
%%%%%%%%%%%%%%%%%%%%%%%%%%%%%%%%%%%%%%%%%%%%%%%%%%%%%%%%%%%%%%%%%%%%%%%%%%%%%%%%%
Whereas in case of ``interaction''-type impurities (first column of \reff{imp_map_vs_B}) a heating of the system by a factor of 2 to $T=0.50$ ($1^\text{st}$ row) leaves only a faint trace of the $\SiSj$ enhancement seen at $T/t=0.25$ ($2^\text{nd}$ row), features caused by ``hopping''-type impurities are much more robust.
In particular, spin correlations over NN-NN bonds which are parallel to imp-imp 
provide (for $t'=0.5t$) strong signals in the large range of elevated temperatures.
Along with this thermal stability 
we also observe another interesting signature: tuning the temperature of the system at constant $t'/t$ 
drives spin correlations between impurity sites from a high-temperature suppression (greenish colors in \reff{imp_map_vs_B}) relative to the impurity-free homogeneous value (black) to a significant low-temperature amplification (reddish colors).
Note that the color scales in \reff{imp_map_vs_B} differ between the rows, reflecting temperature changes of $\SiSj_\text{hom}$ in the impurity-free homogeneous system.

To explore this nontrivial behaviour and its potential use in the cold atom context, let us trace the temperature dependence of the selected spin correlations.
%%%%%%%%%%%%%%%%%%%%%%%%%%%%%%%%%%%%%%%%%%%%%%%%%%%%%%%%%%%%%%%%%%%%%%%%%%%%%%%%%
\begin{figure}[t] %FIG.6%
\includegraphics[width=\columnwidth]{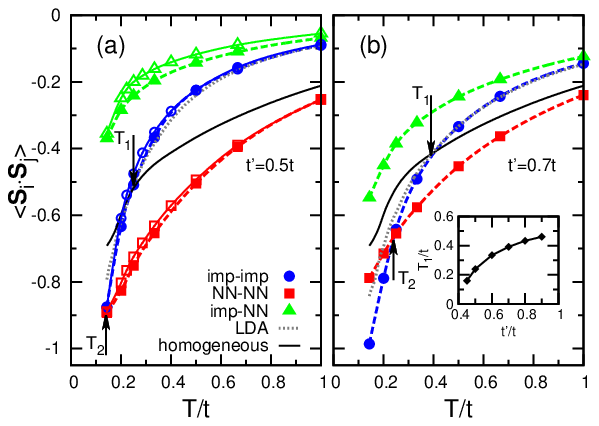}
\caption{(Color online) 
Temperature dependence of NN spin correlation functions $\SiSj$ for a system  with two impurities on  neighboring lattice sites (filled symbols and dashed lines) or with a single line of impurities (open symbols and solid lines). 
$U'=4t=U$, $t_{\text{imp},j}=t'$ with (a) $t'=0.5t$ and (b) $t'=0.7t$.
Bonds are denoted as in \reff{nn_vs_timp}.
Black solid line: $\SiSj_\text{hom}(T)$  for the homogeneous impurity-free system with $U/t=4$.
Gray dotted line: NN spin correlations $\SiSj_\text{LDA}$ for a homogeneous system with the parameters of the impurity (LDA).
Arrows are pointing to the crossing temperatures $T_1$ and $T_2$ discussed in the main text.
Inset: $T_1$ as a function of $t'/t$.
}\label{fig:nn_vs_T}
\end{figure}
%%%%%%%%%%%%%%%%%%%%%%%%%%%%%%%%%%%%%%%%%%%%%%%%%%%%%%%%%%%%%%%%%%%%%%%%%%%%%%%%%
These data are shown in \reff{nn_vs_T} for the three types of bonds discussed above (see also the inset of \reff{nn_vs_timp}).
NN spin correlations in a homogeneous impurity-free system $\SiSj_\text{hom}(T)$ plotted with black solid lines in \reff{nn_vs_T}  correspond to black-encoded values in the colormaps of each row in \reff{imp_map_vs_B}.
We see that imp-imp AF spin corelations are suppressed at high $T$ relatively to the homogeneous impurity-free phase (and relative to the strong NN-NN correlations).
Lowering the temperature causes strong enhancements of correlations between the impurity sites, which then exceed (by absolute value) the homogeneous impurity-free value $\SiSj_\text{hom}(T)$ for $T< T_1$. 
Below some point $T_2< T_1$ imp-imp  spin correlations become even stronger than NN-NN (see arrows in \reff{nn_vs_T}).
This specific form of temperature dependences suggests the use of impurity-induced NN spin correlations as a sensitive thermometer. 
Measuring, e.g., ratios (or differences) of imp-imp and NN-NN (or homogeneous) correlations 
should strongly suppress systematic measurement errors and allow rather precise estimates of $T$.
For obtaining even higher precision, one may adjust the value of $t'/t$ (similar to shifting the counterweight on a steelyard balance) until the selected correlation functions become equal; 
then $T$ can be read off from the curve $T_{1}(t'/t)$ [or $T_{2}(t'/t)$], as plotted in the inset of \reff{nn_vs_T}.

Introducing solitary impurity sites into systems of experimentally relevant sizes is not expected to affect the mean entropy per particle (as well as any other observable mean values) in any noticable way. 
Still, the insertion of impurities will, in general, induce an excess entropy in an otherwise transition-invariant system.
%%%%%%%%%%%%%%%%%%%%%%%%%%%%%%%%%%%%%%%%%%%%%%%%%%%%%%%%%%%%%%%%%%%%%%%%%%%%%%%%%
\begin{figure}[t] %FIG.7%
\includegraphics[width=\columnwidth]{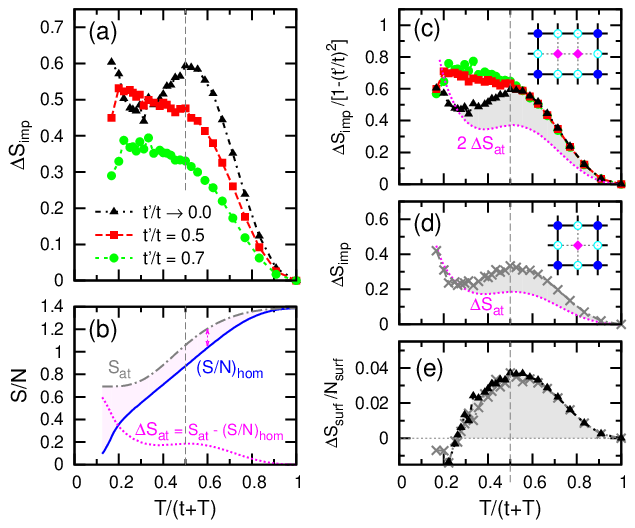}
\caption{(Color online) 
(a) Excess entropy $\Delta S_\text{imp} = S - S_\text{hom}$ of a system with two NN impurities 
[$U'=4t=U$, $t'=0.7t$ (circles), $t'=0.5t$ (squares) or $t'/t \to 0$ (triangles)] versus temperature $T$. 
(b) Entropy $S_\text{at}$ of an isolated ``atomic'' site ($U/t=4, t'/t=0$; dash-dotted line) and
entropy per particle $(S/N)_\text{hom}$ of the reference homogeneous impurity-free system with $U=4t$ (solid line); their difference defines the excess entropy $\Delta S_\text{at}$ per decoupled site (dotted line).
(c) Scaling of the excess entropy [data of panel (a)] with the relative bandwidth change $1-(t'/t)^2$ of each of the two NN impurities; in the limit $t'\to 0$ (black triangles), $\Delta S_\text{imp}$ includes $2\, \Delta S_\text{at}$ (two decoupled atomic sites) plus a surface contribution $\Delta S_\text{surf}$ (shaded; with $N_\text{surf}=6$ surface sites; cf.\ inset).
(d) Excess entropy of single-impurity system at $t'=0$ (crosses); here, $\Delta S_\text{imp}$ includes $\Delta S_\text{at}$ (one decoupled atomic site) plus surface effects $\Delta S_\text{surf}$ (with $N_\text{surf}=4$ surface sites; cf.\ inset).
(e) Surface contribution $\Delta S_\text{surf}$ (per surface site $N_\text{surf}$) to the excess entropy for systems with two NN impurities [triangles, cf.\ (a) and (c)] or a single impurity [crosses, cf.\ (d)]. 
}\label{fig:imp_entropy}
\end{figure}
%%%%%%%%%%%%%%%%%%%%%%%%%%%%%%%%%%%%%%%%%%%%%%%%%%%%%%%%%%%%%%%%%%%%%%%%%%%%%%%%%
\reff{imp_entropy}(a) shows the temperature dependence of 
the excess entropy $\Delta S_\text{imp}$ due to a pair of NN impurities (circles for $t'=0.7t$, squares for $t'=0.5t$ and triangles for the limiting case of fully decoupled impurities $t'/t \to 0$), defined here as a difference in the entropies of equally sized systems with and without impurities. 
These curves demonstrate non-trivial dependencies on both temperature and $t'$. 
The high-temperature tails of $\Delta S_\text{imp}$ for various $t'$ scale perfectly with the relative bandwidth change $1-(t'/t)^2$, 
as demonstrated in \reff{imp_entropy}(c). 
At lower temperatures, at $T\lesssim t$ [i.e., $T/(t+T)\lesssim 0.5$; dashed vertical lines in \reff{imp_entropy}], one observes a significantly reduced slope, until a broad minimum develops at $T/t \sim 0.4$ in the fully decoupled limit $t'/t\to 0$.

As the Hamiltonian can be split up exactly at $t'/t=0$ into $N_{imp}$ contributions describing isolated (``atomic'') impurity sites and the rest of the system (with $N-N_{imp}$) sites, respectively, the same must be true for the associated entropy contributions. 
The dashed-dotted line in \reff{imp_entropy}(b) shows the entropy $S_\text{at}$ of a single isolated interacting site (when $T$ is measured in units of $U/4$), which is seen to exceed the entropy per particle $(S/N)_\text{hom}$ of the reference homogeneous impurity-free system (blue solid line); the difference (shaded region) $\Delta S_\text{at}$ (dotted line) has a local maximum at $T\approx t$ ($= U/4$).
As seen in \reff{imp_entropy}(c), the resulting contribution $2 \Delta S_\text{at}$ (dotted line) accounts for a large part of the total excess entropy of the two-NN-impurity system at $t'=0$ (black triangles); in particular, both agree within precision at $T\lesssim 0.25$. However, a significant difference (shaded) remains at elevated temperatures, with a maximum at $t\approx T$. This second contribution to $\Delta S_\text{imp}$, to be denoted as $\Delta S_\text{surf}$ is clearly associated with the impact of introducing a surface (here by taking $N_\text{imp}=2$ sites out) into an otherwise homogeneous system.

While this impact could, in principle, depend on the exact topology of the surface, we may expect that the primary mechanism is a reduction in the coordination number (from $Z=4$ to $Z=3$) of each of the $N_\text{surf}=6$ surface sites. We should, therefore, expect that $\Delta S_\text{surf}$ is approximately proportional to $N_\text{surf}=6$. 
In order to test this picture, we have also computed the excess entropy of a single-impurity system (at $t'=0$); shown as crosses in \reff{imp_entropy}(d). Again, we can compute the surface contribution $\Delta S_\text{surf}$ (shaded) by subtracting $\Delta S_\text{at}$ of the single involved impurity. Already at first sight, we see that $\Delta S_\text{surf}$ is clearly smaller than in the two-impurity case, as expected for a reduced number $N_\text{surf}=4$ of impurity sites. A closer inspection of the surface contribution per surface site, shown in \reff{imp_entropy}(e) shows that, indeed, $\Delta S_\text{surf}/N_\text{surf}$ agrees, within error bars, between the single-impurity case (crosses) and two-NN-impurity case (triangles) in the accessible temperature range. We conclude that, to a good approximation, the excess entropy in the fully decoupled case $t'/t=0$ is independent of the shape of the boundaries and has contributions linear in the number $\Delta S_\text{imp}$ of impurities and $N_\text{surf}
$ of surface sites, respectively.
In the general situation ($0<t'<t$), all sites remain coupled and, consequently, the total excess entropy will be more complicated and specific to each particular topology.

\vspace{3ex}
%%%%%%%%%%%%%%%%%%%%%%%%%%%%%%%%%%%%%%%%%%%%%%%%%%%%%%%%%%%%%%%%%%%%%%%%%%%%%%%%%%%%%%%%%%%%%%%%%%%%
%%%%%%%%%%%%%%%%%%%%%%%%%%%%%%%%%%%%%%%%%%%%%%%%%%%%%%%%%%%%%%%%%%%%%%%%%%%%%%%%%%%%%%%%%%%%%%%%%%%%
\section{Conclusions}

\vspace{-2ex}
In this study, we considered in detail the effect of localized inhomogeneities on local magnetic phenomena, focussing on nearest-neighbor spin correlations. 
We demonstrated that at or above the spin crossover temperature impurities may induce anomalously large spin correlations in the surrounding. 
These impurity-induced spin correlations possess a non-trivial spatial structure, which cannot be captured within real-space dynamical mean field calculations.
For an impurity type with reduced hopping amplitudes from/to the impurity, we found a special structure of spin correlations that, together with its temperature evolution, suggests to use a pair of such impurities as a sensitive local thermometer in experiments with cold gases on optical lattices.

We thank D. Greif, A. Rapp, D. Rost, and U.\ Schneider for valuable discussions.
Support by the DFG within SFB/TRR 49 is gratefully acknowledged.

%%%%%%%%%%%%%%%%%%%%%%%%%%%%%%%%%%%%%%%%%%%%%%%%%%%%%%%%%%%%%%%%%%%%%%%%%%%%%%%%%%%%%%%%%%%%%%%%%%%%
%%%%%%%%%%%%%%%%%%%%%%%%%%%%%%%%%%%%%%%%%%%%%%%%%%%%%%%%%%%%%%%%%%%%%%%%%%%%%%%%%%%%%%%%%%%%%%%%%%%%


\begin{thebibliography}{16}

\bibitem{Millis2003} See e.g. A. J. Millis, Solid State Commun. {\bf 126}, 3 (2003).
\bibitem{Mackenzie1998} A. P. Mackenzie, R. K. W. Haselwimmer, A. W. Tyler, G. G. Lonzarich, Y. Mori, S. Nishizaki, and Y. Maeno, Phys. Rev. Lett. {\bf 80}, 161 (1998). 
\bibitem{Balastky2006} A. V. Balastky, I. Vekhter, and Jian-Xin Zhu, Rev. Mod. Phys. {\bf 78}, 373 (2006).
% Defects in correlated metals and superconductors http://journals.aps.org/rmp/abstract/10.1103/RevModPhys.81.45
\bibitem{Alloul2009} H. Alloul, J. Bobroff, M. Gabay, and P. J. Hirschfeld, Rev. Mod. Phys. {\bf 81}, 45 (2009).
\bibitem{Sprunger1997} P. T. Sprunger, L. Petersen, E. W. Plummer, E. L{\ae}gsgaard, and F. Besenbacher, Science {\bf 275}, 1764 (1997). 
\bibitem{White2009} M. White, M. Pasienski, D. McKay, S. Q. Zhou, D. Ceperley, and B. DeMarco, Phys. Rev. Lett. {\bf 102}, 055301 (2009).
\bibitem{Modugno2010} G. Modugno, Rep. Prog. Phys. {\bf 73}, 102401 (2010). 
\bibitem{Kondov2013} S. S. Kondov, W. R. McGehee, and B. DeMarco, arXiv:1305.6072.
\bibitem{Schirotzek2009} A. Schirotzek, C.-H. Wu, A. Sommer, and M. W. Zwierlein, Phys. Rev. Lett. {\bf 102}, 230402 (2009).
\bibitem{Nascimbene2009} S. Nascimb\`{e}ne, N. Navon, K. J. Jiang, L. Tarruell, M. Teichmann, J. McKeever, F. Chevy, and C. Salomon, Phys. Rev. Lett. {\bf 103}, 170402 (2009).
\bibitem{Schmid2010} S. Schmid, A. H\"{a}rter, and J. H. Denschlag, Phys. Rev. Lett. {\bf 105}, 133202 (2010).
\bibitem{Zipkes2010} C. Zipkes, S. Palzer, C. Sias, and M. K\"{o}hl, Nature (London) {\bf 464}, 388 (2010).
\bibitem{Bakr2009} W. S. Bakr, J. I. Gillen, A. Peng, S. F\"{o}lling, M. Greiner, Nature {\bf 462}, 74 (2009). 
\bibitem{Weitenberg2011}  C. Weitenberg, M. Endres,	J. F. Sherson, M. Cheneau, P. Schauß, T. Fukuhara, I. Bloch, and S. Kuhr, Nature {\bf 471}, 319–324 (17 March 2011). % NEW Ref
\bibitem{Targonska2010} K. Targo\'{n}ska and K. Sacha, Phys. Rev. A {\bf 82}, 033601 (2010); 
\bibitem{Vernier2011} E. Vernier, D. Pekker, M. W. Zwierlein, and E. Demler, Phys. Rev. A {\bf 83}, 033619 (2011).
\bibitem{Jiang2011} L. Jiang, L. O. Baksmaty, H. Hu, Y. Chen, and H. Pu, Phys. Rev. A {\bf 83}, 061604(R) (2011).
\bibitem{Ohashi2011} Y. Ohashi, Phys. Rev. A  {\bf 83}, 063611 (2011).
\bibitem{Li2012a} J. Li and C. S. Ting, Phys. Rev. B {\bf 85}, 094520 (2012).
\bibitem{Li2012b} J. Li, J. An, and C. S. Ting, Phys. Rev. Lett. {\bf 109}, 196402 (2012).
\bibitem{Jordens2010} R. J\"ordens, L. Tarruell, D. Greif, T. Uehlinger, N. Strohmaier, H. Moritz, T. Esslinger, L. De Leo, C. Kollath, A. Georges,  V. Scarola, L. Pollet, E. Burovski, E. Kozik, and M. Troyer, Phys. Rev. Lett. {\bf 104}, 180401 (2010).
\bibitem{Gorelik2012} E. V. Gorelik, D. Rost, T. Paiva, R. Scalettar, A. Kl\"{u}mper, and N. Bl\"{u}mer, Phys. Rev. A {\bf 85}, 061602 (2012).
\bibitem{Chang2013} C.-C. Chang, R. T. Scalettar, E. V. Gorelik, and N. Bl\"{u}mer, Phys. Rev. B {\bf 88}, 195121 (2013).
\bibitem{Tarruell2012} L. Tarruell, D. Greif, T. Uehlinger, G. Jotzu, and T. Esslinger,  Nature {\bf 483}, 302 (2012).
\bibitem{Paiva2010} T. Paiva, R. Scalettar, M. Randeria, and N. Trivedi, Phys. Rev. Lett. {\bf 104}, 066406 (2010).
\bibitem{Trotzky2010} S. Trotzky, Y.-A. Chen, U. Schnorrberger, P. Cheinet, and I. Bloch, Phys. Rev. Lett. 105, 265303 (2010).
\bibitem{Greif2011} D. Greif, L. Tarruell, T. Uehlinger, R. J\"{o}rdens, and T. Esslinger, Phys. Rev. Lett. {\bf 106}, 145302 (2011).
\bibitem{Greif2013} D. Greif, T. Uehlinger, G. Jotzu, L. Tarruell, and T. Esslinger,  Science {\bf 340}, 1307 (2013).
\bibitem{Esslinger2010} T. Esslinger, Annu. Rev. Condens. Matter Phys. {\bf 1}, 129 (2010). 
\bibitem{Costamagna2006} S. Costamagna, C. J. Gazza, M. E. Torio, and J. A. Riera, Phys. Rev. B {\bf 74}, 195103 (2006). % Hubbard + impurity in d=1, n<1
\bibitem{Wolff1961} P.\ A.\ Wolff, Phys.\ Rev.\ {\bf 124}, 1030 (1961). % Wolff model (1 correlated Hubbard site)
\bibitem{Anderson1961} P. W. Anderson, Phys. Rev. {\bf 124}, 41 (1961). % Anderson model
\bibitem{Kondo1964} J. Kondo, Prog. Theor, Phys. 32, 37 (1964). % Kondo model
% vacancies in Heisenberg model
\bibitem{Martins1997} G. B. Martins, M. Laukamp, J. Riera, and E. Dagotto, Phys. Rev. Lett. {\bf 78}, 3563 (1997).
\bibitem{Song2000} Y.\ Song, H.\ Q.\ Lin, and A.\ W.\ Sandvik, J.\ Phys.: Condens.\ Matter {\bf 12}, 5275 (2000).
\bibitem{Anfuso2006} F. Anfuso and S. Eggert, Europhys.\ Lett.\ {\bf 73}, 271 (2006). % Heisenberg + 1-2 vacancies, various configurations
\bibitem{Engel2009} J. Engel and S. Wessel, Phys. Rev. B {\bf 80}, 094404 (2009). % vacancy + external J´
\bibitem{Metzner1989} W. Metzner and D. Vollhardt, Phys. Rev. Lett. {\bf 62}, 324 (1989). 
\bibitem{Georges1996} A. Georges, G. Kotliar, W. Krauth, and M. J. Rozenberg,  Rev. Mod. Phys. {\bf 68}, 13 (1996).
\bibitem{Blankenbecler1981} R. Blankenbecler, D. J. Scalapino, and R. L. Sugar, Phys. Rev. D {\bf 24}, 2278 (1981). 
\bibitem{QUEST_link} QUEST: QUantum Electron Simulation Toolbox, Version 1.3.0, http://quest.ucdavis.edu.
\bibitem{Rost2012} D. Rost, E.\ V.\ Gorelik, F. Assaad, and N. Bl\"{u}mer, Phys. Rev. B {\bf 86}, 155109 (2012).
\bibitem{Gorelik2010} E. V. Gorelik, I. Titvinidze, W. Hofstetter, M. Snoek, and N. Bl\"{u}mer, Phys. Rev. Lett. {\bf 105}, 065301 (2010).
\bibitem{Jordens2008} R. J\"{o}rdens, N. Strohmaier, K. G\"{u}nter, H. Moritz, and T. Esslinger, Nature {\bf 455}, 204 (2008).
\bibitem{SiSj} $\SiSj = \frac{1}{4}\NNcorr$ (for vectors $\hat{\boldsymbol{\sigma}}_i$ of Pauli matrices).
\bibitem{Helmes2008} R. W. Helmes, T. A. Costi, and A. Rosch, Phys. Rev. Lett. {\bf 100}, 056403 (2008). 
\bibitem{Snoek2008} M. Snoek, I. Titvinidze, C. T\"{o}ke, K. Byczuk, and W. Hofstetter, New J. Phys. {\bf 10}, 093008 (2008).
\bibitem{Gorelik2011} E. V. Gorelik and N. Bl\"{u}mer, Journal of Low Temperature Physics {\bf 165}, 195 (2011).
\bibitem{Bluemer11CCP} N.\ Bl\"umer and E.\ V.\ Gorelik, Comp.\ Phys.\ Comm. {\bf 118}, 115 (2011).
\bibitem{Hirsch86} J.\ E.\  Hirsch and R.\ M.\  Fye, Phys.\ Rev.\ Lett. {\bf 56}, 2521 (1986).
\bibitem{Bluemer07}  N.\ Bl\"umer, Phys. Rev. B {\bf 76}, 205120 (2007).
\bibitem{Bluemer2013} N.\ Bl\"umer and E.\ V.\ Gorelik, Phys.\ Rev.\ B {\bf 87}, 085115 (2013).

 \end{thebibliography}
\end{document}